\documentclass[preprint]{acm_proc_article-sp}
\usepackage{epsfig}
\usepackage[square,comma,numbers,sort&compress]{natbib}
\usepackage{multirow}

\newcommand{\UPLB}{University of the Philippines Los Ba\~{n}os}
\newcommand{\bigO}{\mathcal{O}}
\newcommand{\C}{\mathbb{C}}
\newcommand{\K}{\mathbb{K}}
\newcommand{\Line}{\mathbb{L}}
\newcommand{\Prog}{\mathcal{P}}
\newcommand*\mean[1]{\overline{#1}}

\begin{document}
\title{Synchronization of ad hoc Clock Networks}
\numberofauthors{1}
\author{
\alignauthor Jaderick P. Pabico\\
   \affaddr{Institute of Computer Science}\\
   \affaddr{\UPLB}\\
   \affaddr{College 4031, Laguna}\\
   \email{jppabico@uplb.edu.ph}
}

\date{}
\maketitle
\begin{abstract}
We introduce a graph-theoretic approach to synchronizing clocks in an {\em ad hoc} network of $N$~timepieces. Clocks naturally drift away from being synchronized because of many physical factors. The manual way of clock synchronization suffers from an inherrent propagation of the so called ``clock drift'' due to ``word-of-mouth effect.'' The current standard way of automated clock synchronization is either via radio band transmission of the global clock or via the software-based Network Time Protocol (NTP). Synchronization via radio band transmission suffers from the wave transmission delay, while the client-server-based NTP does not scale to increased number of clients as well as to unforeseen server overload conditions (e.g., flash crowd and time-of-day effects). Further, the trivial running time of NTP for synchronizing an $N$-node network, where each node is a clock and the NTP server follows a single-port communication model, is~$\bigO(N)$. We introduce in this paper a $\bigO(\log N)$ time for synchronizing the clocks in exchange for an increase of $\bigO(N)$ in space complexity, though through creative ``tweaking,'' we later reduced the space requirement to~$\bigO(1)$. Our graph-theoretic protocol assumes that the network is $\K_N$, while the subset of clocks are in an embedded circulant graph $\C_{n<N}^q$ with $q$~jumps and clock information is communicated through circular shifts within the $\C_{n<N}^q$. All $N$~nodes communicate via a single-port duplex channel model. Theoretically, this synchronization protocol allows for $N(\log N)^{-1} - 1$ more synchronizations than the client-server-based one. Empirically through statistically replicated multi-agent-based microsimulation runs, our protocol allows at most 80\% of the clocks synchronized compared to the current protocol which only allows up to 30\% after some steady-state time.
\end{abstract}

\section{Introduction}
The ``Juan Time, On Time'' is a project of the Department of Science and Technology (DOST) launched in 30 September 2011 which aim to campaign for the use of the Philippine Standard Time (PST). Since 1978, the PST is legally and officially maintained by DOST's Philippine Atmospheric, Geophysical and Astronomical Services Administration (PAGASA)~\citep{bp8}. However, due to various reasons, the PST has not been utilized by Filipinos, whether in public or private transactions, resulting to having timepieces that are not synchronized with the PST. There are many problems that result by having non-PST-synchronized timepieces, some possible (though relatively exaggerated) examples of these are:
\begin{enumerate}
\item Historical and official events being recorded with conflicting times -- e.g., in law enforcement, blotters with conflicting records of when crimes were committed may cause the criminal justice system to incarcerate an innocent person or free a guilty one.
\item Financial transactions, specifically those done electronically, may cause one investment to lose a supposedly financial gain -- e.g., an online bidder may submit a bid which might be a second late because her\footnote{Note: The use of the female gender in this paper is just a writing style and this could mean either without being prejudice to the other.} timepiece is not synchronized with that of the bidding institution's.
\item In national defense, the order of a military commander may be executed several seconds earlier or later, instead of on time, which may later prove fatal to national security concerns -- e.g., an air bomber pilot may release a second early a bomb payload to a rebel camp holding up hostages that have not yet been evacuated to a safe zone. 
\item In scientific research that rely on the accuracy and timeliness of the measuring devices -- e.g, a clock-based data monitoring device may provide a sequence of wrong data array because the clock ran faster than expected, which if not corrected may prove crucial to the research conclusion.
\end{enumerate}

\subsection{Clock Drift}

There are many reasons why timepieces are not synchronized with one another, even though they started accurately synchronized. One of the reasons is the ``clock drift''~\citep{murdoch06} which happens because of the following physical reasons:
\begin{enumerate}
\item The clock changed its frequency (i.e., frequency shift).
\item The clock changed its phase (i.e., phase shift).
\item For a limited time (i.e., maybe a burst of several milliseconds), the clock experienced an unstable/interrupted power supply that resulted in either a frequency shift, a phase shift, or both.
\item During an extended use or because of environmental factors, a clock was heated up that resulted in a frequency shift, a phase shift, or both.
\end{enumerate}

\subsection{Word-of-mouth Propagation of Clock Drift}

In the past, and even until now, timepieces are generally updated using the following simple process:
\begin{enumerate}
\item Query a supposedly trusted and authoritative time source, which usually is a person, a radio station announcing a time check, or a TV station showing time; and
\item Manually reset the timepiece to the exact time returned by the time source, without considering the lag time between receiving the information from the source and the time it took to reset the clock.
\end{enumerate}

Because of this process, the recipient of the query answer would have reset her timepiece with an inherent ``clock drift'' due to ``word-of-mouth'' effect illustrated as follows: Given $N$ persons $p_1, p_2, \dots, p_N$, where $p_1$ is an official authoritative source of time. If $p_2$ updates her timepiece by querying $p_1$, and then $p_3$ updates her timepiece by querying $p_2$, and then so on in a linear fashion up to $p_N$ updating her timepiece by querying $p_{N-1}$, at the $(N-1)$th step, $p_N$ would have a clock drift with an optimistic factor of at most $2N$. This factor is due to the ``word-of-mouth'' propagation of the time lag.

\subsection{Clock Synchronization via Radio Transmission}

In advanced countries where timeliness is of utmost importance, like the United States and Japan, timepieces are equipped with (usually an amplitude modulation or AM) radio band receiver~\citep{plagger86} and are updated or synchronized at specified frequency by a signal from a dedicated (usually government-run AM) radio transmitter. The transmitter is connected to a time standard device, such as an atomic clock. Timepieces in these areas automatically adjust to differences in time zones, as well as to changes in daylight saving times (DST). However, timepieces are only adjusted up to a resolution of a second, because the respective AM receivers are not equipped to detect for the propagation delay of the radio signal from the transmitters. On the average, the propagation delay is approximately 1~s for every 300~Km distance the receiver is from the transmitter. Thus, this type of clock synchronization system is effective only to timepieces that only require a resolution of up to a second, which currently are useful for general human use.

\subsection{Internet-based Clock Synchronization via the Network Time Protocol}

The Network Time Protocol (NTP) is a time synchronization protocol implemented in software for the purpose of synchronizing computer clocks over packet-switched, variable latency data networks, such as the Internet. The NTP uses a revised version~\citep{gotoh02} of the Agreement Algorithm, also known as the Marzullo's Algorithm~\citep{marzullo84}, to select time sources for estimating the accurate time from a poll of  noisy sources. Time sources become noisy because of the effects of variable network latency, which the algorithm corrects by using a jitter buffer. The jitter buffer is computed earlier by profiling the round-trip times (RTT) of several zero-payload packets from a source node to a target node in the network. The time is synchronized via a hierarchical, semi-layered system of clock sources, starting from what is termed as Stratum~0, a device that is connected to an atomic clock. Stratum~1 devices are computers that are connected to Stratum~0 devices and normally act as servers for timing requests from Stratum~2 servers. In general, Stratum~$n$ devices connect to Stratum~$n-1$ devices to synchronize time in a hierarchical client-server, master-slave fashion, where the masters are the devices in Stratum~$n-1$ and the slaves are the devices in Stratum~$n$. In the Philippines, no Stratum~0 device has been officially established, even with the launching of DOST's ``Juan Time, On Time'' campaign, which only uses the word-of-mouth propagation of the correct time with up to 1~minute resolution. Despite of this, most computer servers are potential Stratum~1 devices if they connect to known Stratum~0 devices abroad.

\subsection{Potential of Institutions as Statum 1 Service Providers}

Nowadays, various local government and private institutions, particularly those in the highly urbanized areas, run several computer servers for providing ICT services to their constituents~\citep{iglesias10}. Some of these servers might be converted to run in dual-server modes with NTP. A dedicated cluster of NTP servers to act as a publicly-available Stratum~1 devices could be setup but may prove cost ineffective as more client computers connect and query the cluster for correct time at a higher resolution and to synchronize clocks. With the expected improvement of telephone and communication services in the country~\citep{aldaba08}, particularly due to a healthier business competition that the ASEAN integration in 2015 will bring~\citep{hollweg09,balboa10}, it is expected that the use of mobile computers among constituents will double every year. For a relatively small central business district with a pessimistic maximum estimate of 10,000 constituents, each owning at least one mobile computer that query the cluster for correct time, the cluster will be overwhelmed with answering queries for RTTs than for answering queries about the correct time. Thus, it is seen that the NTP is not an efficient protocol for synchronizing the devices beyond Stratum 1~for a very, very large client base.

\subsection{The Solution: Peer-to-peer Protocol for Synchronizing Clocks Beyond Stratum~1}

The problem with using NTP beyond Stratum~1 is that it uses a master-slave type of communication, where the master could be overwhelmed by slaves that number in tens of thousands, especially if the bandwidth does not scale with the increase of estimated users. With a constant bandwidth towards the master, it is necessary that the bandwidth used for answering RTTs and queries be distributed among the participating slaves via what is called a peer-to-peer (P2P) communication approach, similar to the strategy employed by the famous BitTorrent protocol~\citep{pouwelse05,costa08}. Thus, a new protocol is needed to query time and synchronize clocks for devices beyond the Stratum~1 device.

We present in this paper an integrated knowledge in Process Theory and Graph Theory, particularly that of circular-shift process over circulant graphs~$\C$~\citep{whitney32}, to design a protocol for synchronizing $N$~clocks in a complete network $\K_N$ and to show that the $(\log N)$-step protocol is correct and achievable. We show that our clock synchronization protocol is faster by a factor of $\log N$, where~$N$ is the number of timepieces that are concurrently synchronizing.

\section{Improved Berkeley Protocol with Recursive Doubling Technique}\label{sec:3}

In the Berkeley Protocol (BP), given $N$~clocks namely $C_0, C_1, \dots, C_{N-1}$ with time readings $T_0, T_1, \dots, T_{N-1}$, respectively, where $T_0 \ne T_1 \ne \cdots \ne T_{N-1}$, the problem is to synchronize the times without relying on a global clock~$\Gamma$. BP does this by averaging the $N$~time readings with the assumption that no time reading is too extreme to effect a skew to the average. This can be performed in two ways, through an elected leader and through distributed computation. In the first method, an elected leader, usually $C_0$, collects the respective $N-1$ time readings, computes the average $\mean{T}$, and then distributes $\mean{T}$ to $N-1$ others. In the second method, everybody broadcasts their own time readings to others, and they respectively compute the average without anymore additional communication. 

\subsection{The Elected Leader Computes}

\begin{figure}
\centering\epsfig{file=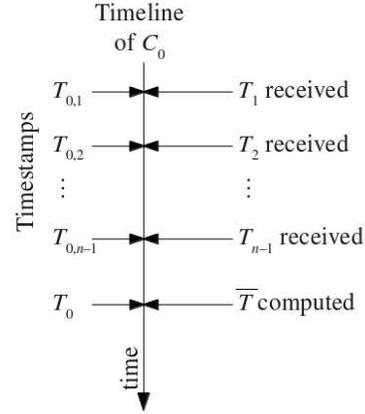, width=2in}
\caption{The timeline of $C_0$ showing when in $C_0$'s own perspective of time it received the respective time readings, as well as when it computed $\mean{T}$.}\label{fig:1}
\end{figure}

In the first method, the collection of the respective time readings takes $N-1$ steps, as the leader $C_0$ needs to retrieve the time readings of $C_1, C_2, \dots,$ and $C_{N-1}$ one at a time. To compensate for the elapsed time due to collection of each time readings, every time a reading $T_i$ is received, $C_0$ puts its own timestamp $T_{0,i}$ on it. At the end of the $N-1$ collection steps, $C_0$ would have collected the time readings $T_1, T_2, \dots,$ and $T_{N-1}$ with respective timestamps $T_{0,1}, T_{0,2}, \dots,$ and $T_{0,N-1}$. At the time of computation, which interestingly is at $T_0$, the $i$th time reading would have aged $T_0 - T_{0,i}$, thus $T_i$ must be corrected with this difference. Figure~\ref{fig:1} shows the timeline of $C_0$ with respect to the receipt of the time readings at the respective timestamps. 

\subsubsection{Computation of the $\mean{T}$}

The average time $\mean{T}$ is computed as follows:

\begin{eqnarray}
	\mean{T} &=&	\frac{1}{N}\bigg(T_0 + T_1+(T_0-T_{0,1}) + \bigg.\nonumber\\
	        & & \qquad\qquad T_2+(T_0-T_{0,2}) + \dots + \nonumber\\
	        & & \qquad\qquad \bigg.T_{N-1}+(T_0-T_{N,N-1}) \bigg)\nonumber\\
            &=& \frac{1}{N}\bigg( T_0 + T_0+(T_1-T_{0,1}) + \bigg.\nonumber\\
	        & & \qquad\qquad  T_0+(T_2-T_{0,2}) + \dots + \nonumber\\
            & & \qquad\qquad \bigg. T_0+(T_{N-1} - T_{0,N-1})\bigg)\nonumber\\
    \mean{T} &=& \frac{1}{N}\left(N T_0 + \sum_{i=1}^{N-1} \left( T_i - T_{0,i}\right)\right)\label{eqn:1}\\
    \mean{T} &=& \frac{1}{N}\left(N T_0 + \sum_{i=1}^{N-1} T_i - \sum_{i=1}^{N-1} T_{0,i}\right)\label{eqn:2}
\end{eqnarray}

In the above equations, it would have sufficed to stop with Equation~\ref{eqn:1} but we will soon see that the form in Equation~\ref{eqn:2} is practically useful in optimizing the space complexity of the methodology. The space complexity requirement of this method is discussed further below (Subsection~\ref{sec:3.1.3}).

It would have taken $T_{0,c}$ time to compute for $\mean{T}$, thus $\mean{T}$ must be corrected with this amount of computation time also. After correction, $\mean{T} + T_{0,c}$ will be distributed by $C_0$ to the $N-1$ other clocks. This will be done by $C_0$ one clock at a time for a total of $N-1$ steps, where each step, the elapsed time due to the previous communication will be added to the corrected $\mean{T}$. Thus, $C_1$ will receive $\mean{T} + T_{0,c}$,  $C_2$ will receive  $\mean{T} + T_{0,c} + D_{0,1}$, where $D_{0,1}$ is the elapsed time when $C_0$ sent the new time reading to $C_1$,  $C_3$ will receive $\mean{T} + T_{0,c} + D_{0,1} + D_{0,2}$, where $D_{0,2}$ is the elapsed time when $C_0$ sent the new time reading to $C_2$, and so on. In general, the $i$th clock will receive $\mean{T} + T_{0,c} + \sum_{j=2}^{i-1} D_{0,j}$, $\forall 1 < i < N$. 

\subsubsection{Time Complexity Requirement}\label{sec:3.1.2}
This method takes $N-1$ steps to collect the respective time readings, one step to compute for the average, and $N-1$ steps to distribute the corrected average for a total of $2N-2$ steps. Thus the time complexity of this method is $\bigO(N)$. 

\subsubsection{Space Complexity Requirement}\label{sec:3.1.3}
Intuitively, $C_0$ needs $N-1$ spaces to hold the $N-1$ collected time readings. This is what Equation~\ref{eqn:1} provides at a glance. However, $C_0$ can just use 2 spaces to separately hold the running sum of the collected time readings and the running sum of the timestamps. This is what Equation~\ref{eqn:2} is showing. $C_0$ can reuse one of the two spaces to hold the corrected $\mean{T}$. Thus, this method's best space complexity is $\bigO(1)$. 

\subsubsection{The Pitfall of Simplicity}
\begin{figure}
\centering\epsfig{file=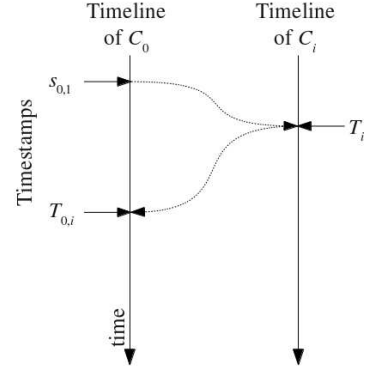, width=2in}
\caption{The respective timelines of $C_0$ and $C_i$ showing the time values elapsed when initiating at timestamp $s_{0,i}$ and completing at timestamp $T_{0,i}$ the collection of $T_i$ from $C-i$.}\label{fig:2}
\end{figure}

Regardless of the time and space complexities, the method suffers from simplicity because it did not consider the additional time it will take for the time readings to reach $C_0$ from their respective clocks. In Figure~\ref{fig:1} above, $T_i$ is basically the same as $T_{0,i}$, $\forall 0 < i < N$. This is not the case, however, because each clock either runs faster or slower than $C_0$. When $C_0$ collects data from $C_i$, it must have recorded the timestamp $s_{0,i}$ at the start of its communication with $C_i$. Upon receipt of the time reading $T_i$ from $C_i$, $C_0$ must have also recorded the timestamp $T_{0,i}$. If $C_i$ is synchronized with $C_0$, definitely $s_{0,i} < T_i < T_{0,i}$. If we assume that the time it takes for a request from $C_0$ to reach $C_i$ is the same as the time it takes for the response from $C_i$ to reach $C_0$, then that time is $E_{0,i} = 0.5 (s_{0,i} + T_{0,i})$. The amount $(s_{0,i} + T_{0,i})$ is known in the literature as the roundtrip time (RTT)~\citep{biaz03,sessini06,arreyouchi13}, and therefore ${\rm RTT}_{0,i} = 2  E_{0,i}$. This amount is the one missing in the above discussion. Figure~\ref{fig:2} shows the visualization of these time values between the exchange of $C_0$ and $C_i$.

Considering the asynchronous nature of the clocks, we can now obtain an estimate for $T_i$ that is closer to its correct value and it is given as $T_i + 0.5 {\rm RTT}_{0,i}$. With this corrected value, Equation~\ref{eqn:2} must also be corrected into:
\begin{eqnarray}
	\mean{T}&=&\frac{1}{N}\left(N T_0 + \sum_{i=1}^{N-1} T_i + \right.\nonumber\\
	        && \qquad \left. \frac{1}{2} \sum_{i=1}^{N-1} {\rm RTT}_{0,i} - \sum_{i=1}^{N-1} T_{0,i}\right)
\end{eqnarray}

\subsubsection{Computation of the {\rm RTT}}

Where will the RTT's come from? Here, we propose a methodology that minimizes the error of the estimate for RTT. The reason for the error is that the time it will take for $C_0$'s request to reach $C_i$ is almost always not the same as the time it will take for $C_i$'s response to reach~$C_0$. Depending on the implementation of the communication protocol, $C_0$'s initial request might as well go as little as one bit in length, say the value 0 upon receipt by $C_i$ to mean that the elected leader, $C_0$, is requesting~$C_i$ to send its time reading $T_i$. The response, however, could involve a 32-bit integer, representing the number of seconds since some reference year. The propagation of a 1-bit data is faster than the propagation of a 32-bit data, especially to bandwidth constrained communication channels. Thus, we want to create a methodology that ensures that the RTT is relatively constant during the time of the collection of the time readings, and at the same time, we want to read $T_i$ while this RTT is seemed to be non-changing.

We propose the following algorithm:

{\bf Algorithm~1:} Computation of ${\rm RTT}_{0,i}$ with $T_i$
\begin{enumerate}
\item Set $j = 0$.\label{step:1}
\item Repeat the following:\label{step:2}
  \begin{enumerate}
  \item Increment $j$ by 1
  \item $C_0$ sends a 1 to $C_i$ at time $s_{0,i}$
  \item $C_0$ receives a 32-bit long data from $C_i$ at time $T_{0,i}$
  \item $C_0$ computes for the ${\rm RTT}_{0,i,j} = T_{0,i}-s_{0,i}$
  \end{enumerate}
	until $j =$ some statistically possible value
\item Compute for the average $\mean{{\rm RTT}_{0,i}^A} = j^{-1} \sum_{k=1}^j {\rm RTT}_{0,i,k}$ and its standard deviation $\sigma_{0,i}^A$.
\item If $\sigma_{0,i}^A$ is within some set allowed threshold, then we move to step~\ref{step:5}, else we go back to step~\ref{step:1}.
\item $C_0$ sends a 0 to $C_i$ at time $s_{0,i}$\label{step:5}
\item $C_0$ receives the 32-bit long $T_i$ from $C_i$ at time $T_{0,i}$
\item We set $j=0$ and repeat the steps in~\ref{step:2} to collect $j$~${\rm RTT}_{0,i,j}$'s.
\item Compute for the average $\mean{{\rm RTT}_{0,i}^B} = j-1 \sum_{k=1}^j {\rm RTT}_{0,i,k}$ and its standard deviation $\sigma_{0,i}^B$.
\item If $| \mean{{\rm RTT}_{0,i}^A}-\mean{{\rm RTT}_{0,i}^B} | <$ some threshold and $|\sigma_{0,i}^A - \sigma_{0,i}^B| <$ some threshold,
\begin{itemize}
\item then $C_0$ accepts $T_i$ with $\mean{{\rm RTT}_{0,i}^A}$, 
\item else we repeat the whole process from step~\ref{step:1}.
\end{itemize}
\end{enumerate}

We want to set~$j$ in Algorithm~1 such that the time it takes to compute for the $\mean{RTT}$ will not dominate the time it takes to exchange the respective~$T$'s. Unfortunately, $j$~will depend on the state of the underlying network which can only be set through experience. We assume, however, that the network will not be a factor and that we can set~$j$ to a value that can provide a statistically acceptable degree of freedom. We then further assume that the contribution of this algorithm to both the leader computes and the distributed computation approaches is~$\bigO(1)$.

\subsubsection{Improvement of the Steps in Collecting Time Readings}
The collection of time readings in the original protocol, as shown in subsection~\ref{sec:3.1.2}, takes $N-1$ steps, or a time complexity of $\bigO(N)$. We improved this time complexity to $\bigO(\log N)$ by utilizing a recursive doubling technique which we illustrate here with $N = 8$ as follows. The procedure is completed in 3 steps, instead of seven steps. At step 1, $C_0$ sends a 0 to $C_4$. The 0 bit sent by $C_0$ will be propagated first to all clocks, while clocks which have already received the bit will participate in sending. At step 2, $C_0$ sends a 0 to $C_2$, while $C_4$ propagates the 0 to $C_6$. At step 3, $C_0$ sends a 0 to $C_1$, while at the same time $C_2$ propagates the 0 to $C_3$, $C_4$ to $C_5$, and $C_6$ to $C_7$. After step 3, all clocks would have received the 0 from $C_0$. 

The sending of the respective time readings will be done in the opposite manner, also in three steps as follows: At step 1, $C_0$ receives $T_1$ from $C_1$, and at the same time, $C_2$ receives $T_3$ from $C_3$, $C_4$ receives $T_5$ from $C_5$, and $C_6$ receives $T_7$ from $C_7$. All pairs will follow the procedure outlined in Algorithm~1. At step~2, $C_0$ receives $T_2$ and the corrected $T_3$ from $C_2$, while $C_4$ receives $T_6$ and the corrected $T_7$ from $C_6$, again both utilizing Algorithm~1. At step 3, using Algorithm~1, $C_0$ receives $T_4$, $T_5$, $T_6$, and $T_7$ from $C_4$. 

In general, time readings are collected by $C_0$ via a recursive doubling method in $\bigO(\log N)$ steps. However, the space complexity has increased to a corresponding $\bigO(\log N)$ from $\bigO(1)$. Notice that the amount of data being transferred from $C_i$ to $C_0$ doubles every step. Since the total number of doubling is also $\log N$ for $N$~clocks, then the maximum amount of data to be passed is $\log N$ times of the original one. This maximum happens in the last step, though.

\subsubsection{Distributing $\mean{T}$ in $\bigO(\log N)$ Time}
After $C_0$ has computed the $\mean{T}$, it will distribute the average time to $N-1$ clocks via the same recursive doubling technique. The corresponding time complexity is $\bigO(\log N)$ while the space complexity is $\bigO(1)$.

\subsection{Distributed Computation of $\mean{T}$}\label{sec:3.2}
In the second method, each of the clocks $C_0, C_1, \dots,$ and $C_{N-1}$ will collect time readings $T_0, T_1, \dots$, and $T_{N-1}$ from the respective other clocks. Once this collection is completed, each of the clock will perform the averaging on their own, without any more further communication to the other clocks. Thus, our analysis focuses on a particular distribution scheme for the time readings. Intuitively, each clock can perform a collection of time readings from other clocks, one at a time. That is, each clock will be elected as a leader, collect the time readings, and then compute the average for itself without sharing it. If this is done in lexicographic way, and since we have already shown earlier in Subsection~\ref{sec:3.1.2} that this particular method takes $\bigO(N)$ time complexity, then this method, intuitively will cost $\bigO(N \log N)$ time. The question to be asked, then, is can we do better than this?

\subsubsection{The Circular Shift Operation}
\begin{figure}
\centering\epsfig{file=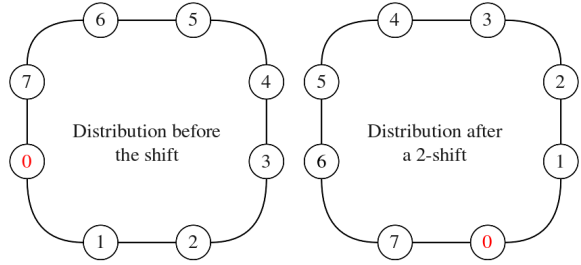, width=3in}
\caption{An example circular 2-shift on a $\C_8^1$, which can be done via a series of two circular 1-shift operations.}\label{fig:3}
\end{figure}

Given a set of $N$~nodes $V_1, V_2, \dots, V_N$ that form a regular circulant graph of order~$N$ with $q$~jumps (or simply~$\C_N^q$)~\citep{whitney32,pabico2014}, the circular $q$-shift operation~\citep{oshiba72,maslov73,gruber09} is a special permutation of the nodes' indexes such that node $V_i$ sends a data packet to node $V_{(i + q) \mod N}$ (Figure~\ref{fig:3}). Researchers have long proved that the optimal number of steps for a circular $q$-shift on a~$\C_N^q$ is $\min(q, N-q)$. To improve the performance of the distribution methodology discussed in Section~\ref{sec:3.2} above, we have to assume that the clocks are arranged in a $\C_N^1$. This is not impossible to do since any~$\C_N^q$ will perfectly embed into a~$\K_N$~\citep{mutzel00}.

Intuitively, the distribution of the time readings to all clocks only needs a circular $(N-1)$-shift operation, which only requires $\min\big(N-1, N-(N-1)\big) = 1$ operation. One can argue that this is true because a circular $(N-1)$-shift operation is equivalently a circular $(-1)$-shift operation (i.e., a circular 1-shift operation in the opposite direction). This is not the case, however, as we will soon see in our modification to the circular $q$-shift operation discussed in the next subsection.

\subsubsection{Circular $(N-1)$-Shift-Copy Operation}

We use the fact that a circular $(N-1)$-shift operation can be done by a series of $(N-1)$ circular 1-shift operations. We modified, however, each circular 1-shift operation such that the receiving clock copies the time readings that has been shifted to it. We call our new operation as a Circular $q$-Shift-Copy Operation. The circular $(N-1)$-shift-copy operation is simply a series of $N-1$ alternating circular 1-shift and copy operations. 

If a copy operation takes 1 step, then our circular $q$-shift-copy operation takes $2(N-1)$ steps, or a complexity of $\bigO(N)$, a vast improvement to the intuitive time complexity discussed in Section~\ref{sec:3.2}, which is $\bigO(N \log N)$. Since each circular 1-shift-copy operation only requires sending 1 data item per operation, then the circular 1-shift-copy operation takes a space complexity of $\bigO(1)$. However, the receiving node must allocate a buffer that is equal to the amount of data that will be shifted, so the operation can take a maximum space complexity of $\bigO(N)$. We can strategically reduce this space complexity to $\bigO(1)$ if for every intermediate circular 1-shift-copy operation, the sum of the copied time readings will already be computed.

\begin{figure}
\centering\epsfig{file=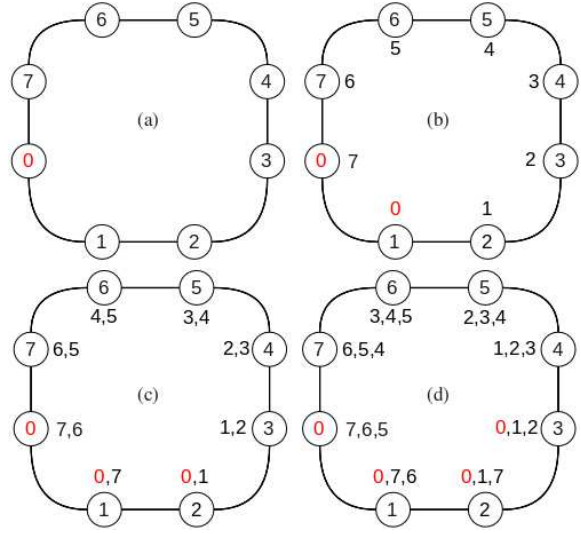, width=3in}
\caption{An example progression of a circular 7-shift-copy operation on a $\C_8^1$: (a) The data distribution before the circular 7-shift-copy operation; (b) The data distribution after the first 1-shift-copy operation; (c) The data distribution after the second 1-shift-copy operation; and (d) The data distribution after the third 1-shift-copy operation.}\label{fig:4}
\end{figure}

Figure~\ref{fig:4} shows a visualization of the progression of the first three 1-shift-copy operations on an $N = 8$ clock network.

\subsubsection{Circular $(N-1)$-Shift-Copy Operation with Recursive Doubling}

The circular $(N-1)$-shift-copy vastly improves the time complexity of the operation from $\bigO(N \log N)$ down to $\bigO(N)$. The next question is, can we do better? It turns out that the answer to the question is a resounding yes as we shall soon see with our new proposed method we called recursively-doubled circular $(N-1)$-shift-copy. This method takes the time complexity of $\bigO(\log N)$ steps, which we will describe as follows:
\begin{enumerate}
\item During the first step, instead of assuming that the clocks were arranged in a $\C_N^1$, we assumed that the clocks were arranged in a $\C_N^{\lfloor N/2 \rfloor}$. This means that clock $C_i$ will be connected to clocks $C_{i+\lfloor N/2\rfloor}$ and $C_{i-\lfloor N/2\rfloor}$, $\forall 0\le i < N$. Such a circulant graph contains $\lfloor N/2\rfloor$ disconnected $\C_2^1$'s. These subgraphs can alternately be seen as a linear graph $\Line_2$ of order~2. The circular 1-shift-copy operation can be performed in these subgraphs concurrently.
\item During the second step, we assumed that the clocks were arranged in a $\C_N^{\lfloor N/4 \rfloor}$, where each clock $C_i$ will be connected to clocks $C_{i+\lfloor N/4\rfloor}$ and $C_{i-\lfloor N/4\rfloor}$, $\forall 0\le i < N$. Such a circulant graph contains $\lfloor N/4\rfloor$ disconnected $\C_4^1$. As in the previous step, these subgraphs can concurrently perform a circular 1-shift-copy operation each. In general, at step~$k$,  we assumed that the clocks were arranged in a $\C_N^{\lfloor N2^{-k} \rfloor}$. The original network will be composed of $\lfloor N2^{-k}\rfloor$ disconnected $\C_{2^k}^1$'s. These subgraphs will concurrently perform a circular 1-shift-copy operation each to distribute the data.
\item At the last step (i.e., $(\log N)$th step), the clock will be assumed to be arranged in a $\C_N^1$, where the circular 1-shift-copy operation distributes the final set of time readings.
\end{enumerate}

\begin{figure}
\centering\epsfig{file=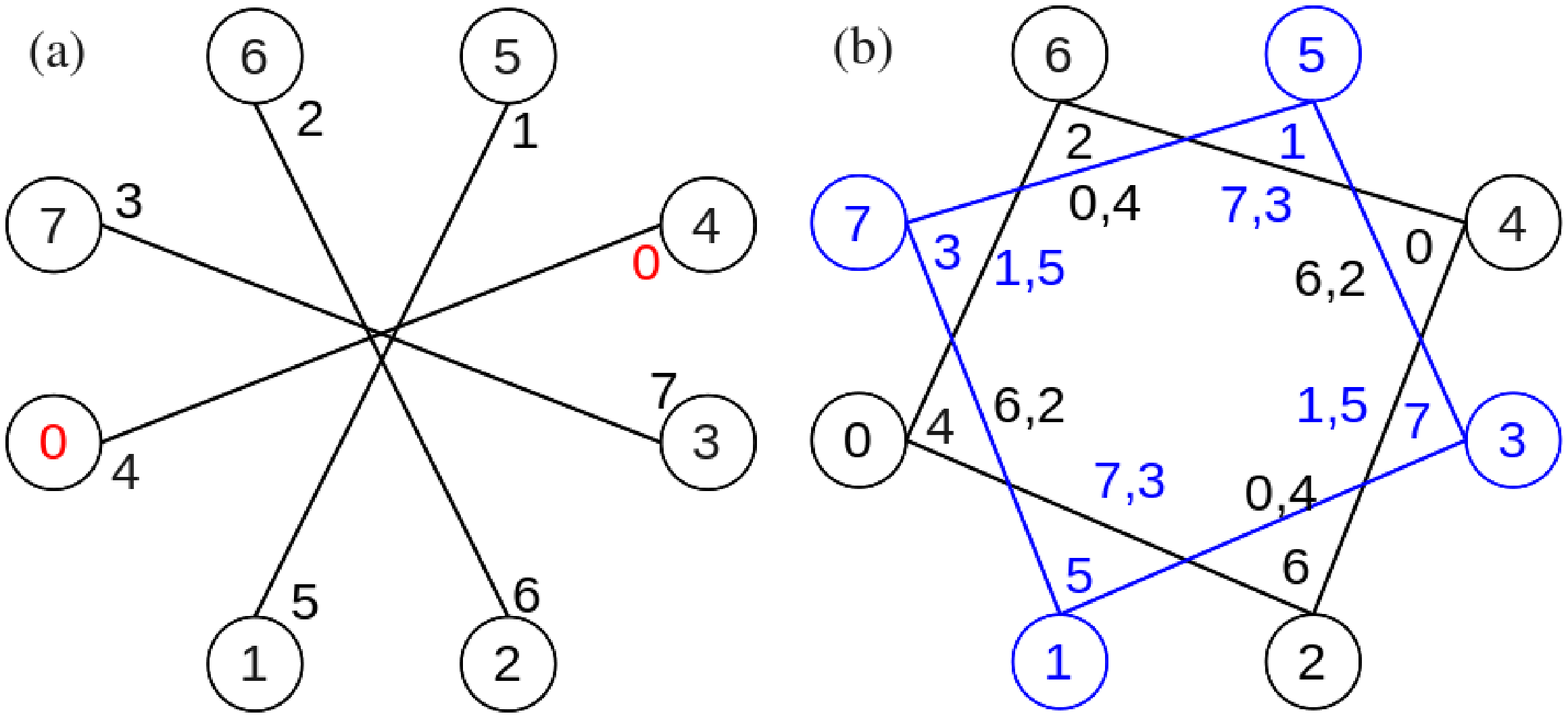, width=3in}
\caption{An example progression of a recursively doubled circular 7-shift-copy operation on a $\C_8^1$: (a) The data distribution after step~1 where the clocks were arranged as a $\C_8^4$; and (b) The data distribution after step~2 where the clocks were arranged as a $\C_8^2$.}\label{fig:5}
\end{figure}

In this new method, the distribution of the time readings takes a time complexity of $\bigO(\log(N))$. Figure~\ref{fig:5} shows the evolution of the circulant graphs at each step of the methodology with $N = 8$.

\subsubsection{Complexities of the Recursively Doubled Circular $(N-1)$-Shift-Copy}
Intuitively, the time complexity of the recursively doubled circular $(N-1)$-shift-copy operation is $\bigO(log N)$. Notice however that the space complexity doubles every step, with the $(\log N)$th step taking $N$~spaces. Obviously, the space complexity is $\bigO(N)$. However, we can strategically reduce the space complexity if at every step we already compute the sum of the time readings. Thus, the space complexity of our proposed method can be as good as $\bigO(1)$ space.
\pagebreak
\subsection{Comparison between Leader Computes and Distributed Computation}


Both the Elected Leader Computes and the Distributed Computation have the same time and space complexities of $\bigO(\log N)$ and $\bigO(1)$, respectively. This does not mean that we can now select any of the two in the implementation of the methodology. For all practical purposes, we chose to implement the distributed computation method because the elected leader computes method will suffer from being ``orphaned'' when the elected leader decided to leave the network at the middle of the computation. Thus, we see the distributed computation method as a more robust method from the dynamism brought about by the constant movement of the clocks in and out of the {\em ad hoc} network.


\subsection{Implementation of the Distributed Computation through Computer Network Simulation}\label{sec:lan}

We implemented the distributed computation by writing a program $\Prog$ that averages the internal clocks of computers connected in a local area network (LAN). We used a simple socket programming~\citep{stevens98} so that clock information can be distributed among the computers in the LAN, using the efficient recursively doubled circular $(N-1)$-shift-copy operation. 

\begin{figure*}
\centering\epsfig{file=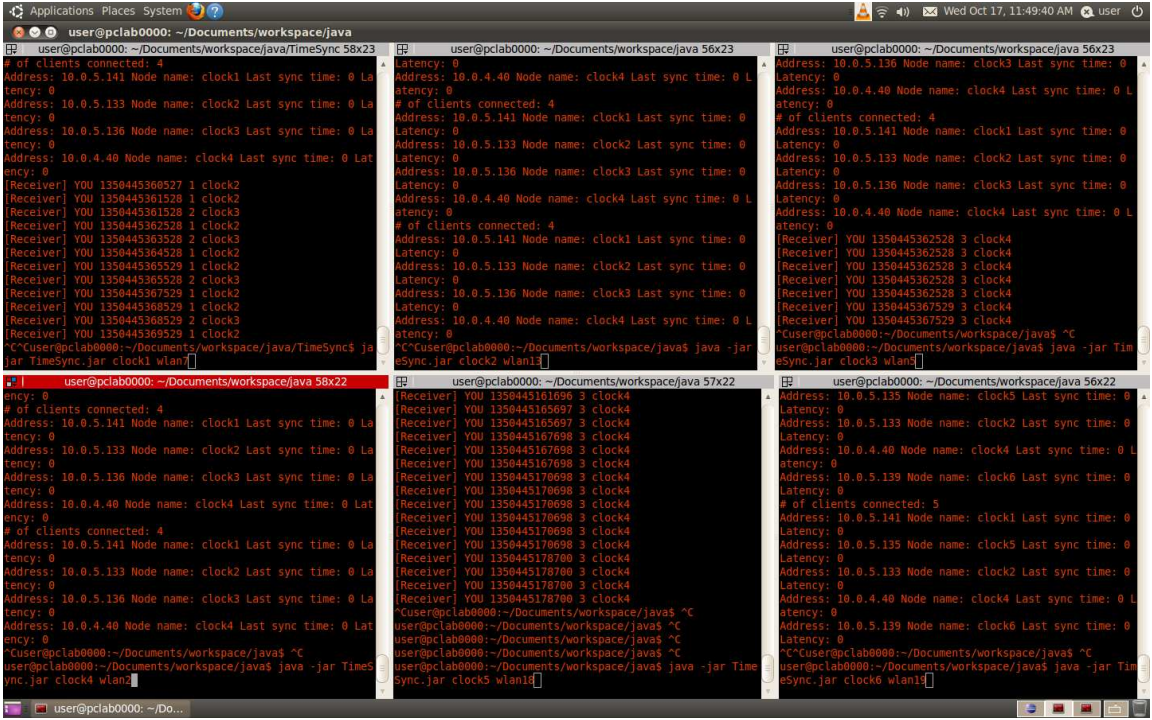, width=6in}
\caption{Screen capture of executing the clock synchronization application $\Prog$ running on a 6-computer LAN through a remote secured shell (SSH) session. Shown in this screen capture are six terminals, each connected to the different computers where internal clock of each is being synchronized.}\label{fig:6}
\end{figure*}

Figure~\ref{fig:6} shows the screen capture of an $(N = 6)$-clock synchronization problem implemented in six LAN-connected x686 processors, each running a multi-programming Gnu/Linux operating system.

\section{Time Synchronization with Multi-agent System}\label{sec:mas}

Our multi-agent-based~\citep{niazi11,salamon11} time synchronization protocol is basically composed of three simple steps that can be implemented by any mobile simulated clock $C_i$ depending on what state $C_i$ is in: (1) Update own time from the global clock $\Gamma$; (2) Update of own time from other clocks, and (3) Update other clocks. In our protocol, we assumed  that each clock has the following:
\begin{enumerate}
\item Time Record ($T_i$) -- The current time of each clock; each clock has different time records since they are not synchronized;
\item $\Gamma$ Synchronization Record ($\Gamma{\rm SR}_i$) -- A time record when clock $C_i$ last synchronized its time from $\Gamma$; and
\item Clock Synchronization Record (${\rm CSR}_i$) -- A time record when $C_i$ last synchronized its time from other clocks.
\end{enumerate}
The clock $C_i$ can be in any of these two states:
\begin{enumerate}
\item ${\rm IN}(\Gamma)$ -- This means that $C_i$ is under the influence of a global clock $\Gamma$; and
\item ${\rm OUT}(\Gamma)$ -- This means that $C_i$ is not under the influence of $\Gamma$.
\end{enumerate}

\subsection{Time Synchronization under ${\rm IN}(\Gamma)$}

When $C_i$ is under the influence of a global clock $\Gamma$, $C_i$ immediately synchronizes with $\Gamma$ via a peer-to-peer protocol (P2P). The immediacy of the synchronization scheme assures those clocks which enter the circle of influence in an almost tangent to the edge of synchronization. Entering at a tangent means that these clocks will soon be out of the influence of $\Gamma$. Whenever $C_i$ synchronizes with $\Gamma$, it updates with it its $\Gamma{\rm SR}$ as well. After the first synchronization, $C_i$ may either be in one of the two available modes: (1) Passive Mode; or (2) Aggressive Mode. These modes were developed to favor those clocks which are equipped with ranging-capable device.  A clock with no ranging capability automatically chooses the aggressive mode, while a clock with ranging capability first chooses the passive mode and then switches to aggressive mode. When a range-capable clock $C_i$ can range that its distance from $\Gamma$ is decreasing, it uses the passive mode. However, when $C_i$ can sense that its distance from $\Gamma$ is increasing, then it switches to the aggressive mode. The passive mode allows for the conservation of power, especially for those clocks  that are powered by batteries.

\subsubsection{Passive Mode at ${\rm IN}(\Gamma)$ State}

Upon entry of $C_i$ into the influence of~$\Gamma$, it first queries the~$\Gamma$ which always returns the current global time~$G$. If $|G-T_i|$ is lesser than some threshold value $Th$, then $C_i$ does not do anything. However, the moment $|G-T_i|> Th$, $C_i$ immediately updates its $T_i$ with $G$, as well as its $\Gamma{\rm SR}$. While still under the influence of~$\Gamma$, $C_i$ continually queries the $\Gamma$ for $G$, until $|G - T_i|>Th$.

\subsubsection{Aggressive Mode at ${\rm IN}(\Gamma)$ State}

Regardless of the $|G-T_i|$ compared to $Th$, $C_i$ always immediately updates its $T_i$ with~$G$. The aggressive mode assures the clock that it always has the most recent $\Gamma{\rm SR}$ upon leaving the influence of~$\Gamma$.

\subsection{Time Synchronization under ${\rm OUT}(\Gamma)$}

\begin{figure}
\centering\epsfig{file=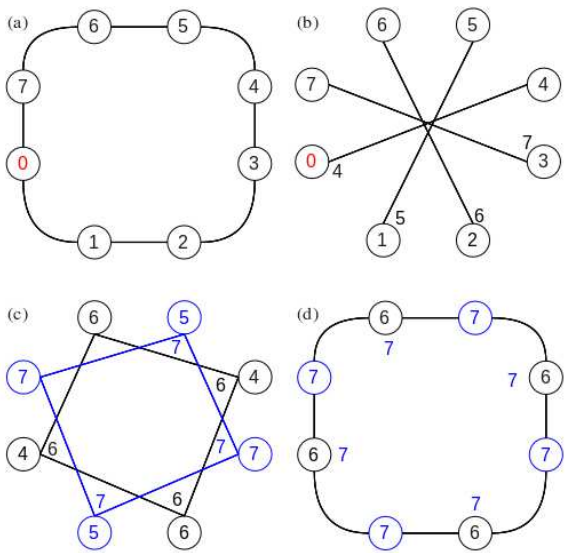, width=3in}
\caption{An example progression of a recursively doubled circular 7-shift-max operation on a $\C_8^1$: (a) The data distribution before step 1; The respective data distributions after steps 1 (b), 2 (c), and 3 (d).}\label{fig:7}
\end{figure}

When a clock $C_i$ is out of the influence of~$\Gamma$, then it could be under the influence of other clocks $C_j$, $\forall j\ne i$ within its immediate broadcast vicinity. Assuming that $C_i$ enters a broadcast vicinity of $N-1$ other clocks, then we can use the protocols discussed in Section~\ref{sec:3}, particularly the distributed computation scheme. However, we will modify the protocol to compute for the $\max_{i=0}^{N-1}(\Gamma {\rm SR}_i)$ instead. We now propose a new method we called recursively doubling circular $(N-1)$-shift-max operation, where at each step of the operation, $C_i$ compares its $\Gamma {\rm SR}_i$ with what it received from its immediate neighbor, and retains the maximum between the two. This operation runs in $\bigO(\log N)$ time complexity and, since we only need to get the maximum, definitely with $\bigO(1)$ space complexity. Figure~\ref{fig:7} shows the visualization of the progression of time synchronization of $N$~clocks using the recursively doubling $(N-1)$-shift-max method.

For clocks with ranging capabilities, they will select to include those clocks that approach them into the network to prolong the life of their {\em ad hoc} community. Definitely, those clocks that are already going away from them will soon be out of their group's circle of influence. We do not want to include those clocks which may leave the network before the synchronization is completed. 

\subsection{Simulation of the Protocol}

\begin{figure}
\centering\epsfig{file=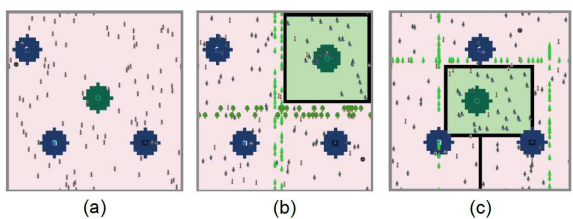, width=3in}
\caption{Snap shots of the multi-agent implementation of the second protocol using a simulation environment: (a)~Scenario~A; (b)~Scenario~B; and (3)~Scenario~C.}\label{fig:8}
\end{figure}

This protocol was simulated using a multi-agent-based simulation environment~\citep{kornhauser07,castro13a,castro13b,muscalagiu13}. We considered three scenarios as follows (Please refer to Figure~\ref{fig:8}):
\begin{enumerate}
\item Scenario~A -- In this scenario, we located the global clock (green circle) at the middle of the environment, and placed three synchronization-disrupting areas (blue circles). Clocks are symbolized by the person icons, which randomly roam about the environment. 
\item Scenario~B -- This scenario is similar to Scenario~A with the difference that the global clock is inside a fenced area and only those authorized persons are allowed to enter the area. This simulates the situation wherein the global clock is only available to a few select people and that time synchronization will only happen if these select people will come in contact with those that were not selected.
\item Scenario~C -- This scenario is similar to Scenario~B but this time the fenced global clock is already located at the center, while the synchronization-disrupting areas are placed near the fence of the global clock.
\end{enumerate}

\begin{figure}
\centering\epsfig{file=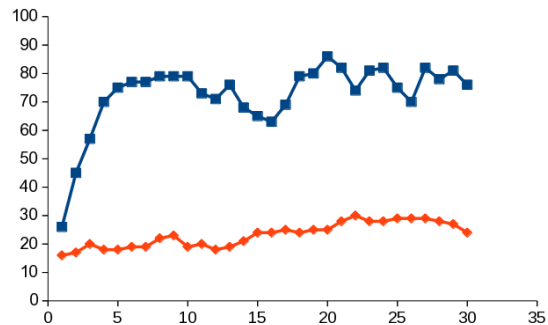, width=3in}
\centering\epsfig{file=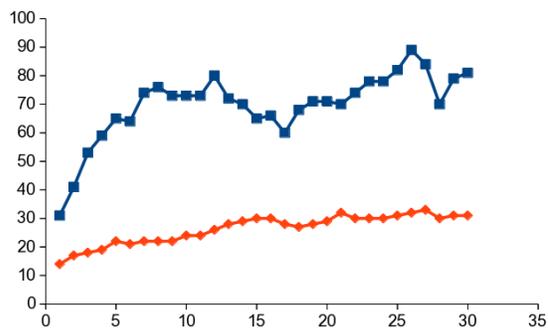, width=3in}
\centering\epsfig{file=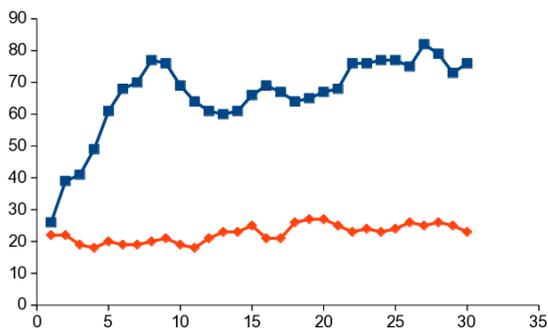, width=3in}
\caption{Plot of the percentage of $\Gamma$-synchronized clocks during the first 30-s of the simulation for Scenario~A (top line plot), Scenario~B (middle line plot), and Scenario~C (bottom line plot). The horizontal axis is in seconds while the vertical axis is in percentage of $\Gamma$-synchronized clocks. Blue lines with square points are for the proposed protocol while orange lines with diamond points are for the simple protocol.}\label{fig:9}
\end{figure}

For comparison purposes for each scenario, we implemented a simple protocol that mimics how the current time-synchronization is currently being implemented. Synchronization happens when a newly $\Gamma$-synchronized clock $C_x$ meets another clock $C_i$. In this protocol, $C_x$ always shares its time with $C_i$ via a simple P2P data exchange. Figure~\ref{fig:9} shows the percentage of $\Gamma$-synchronized clocks within the first 30-s of the simulation time. This figure shows that the synchronization protocol that we developed can provide about 70\% to 80\% synchronous clocks while the simple protocol can only provide up to 30\% synchronous clocks for any scenario.

\pagebreak
\section{Conclusion}
In this paper, we argued that the DOST's  ``Juan Time, On Time'' program of using the PST with a simple synchronization protocol does not provide high percentage of $\Gamma$-synchronized clocks because of the inherrent clock drift brought about by the simple protocol. In fact, the clock drift is even enhanced by the simple protocol. We then provide an alternative automated protocol that synchronizes $N$~clocks in $\bigO(\log N)$ time using only $\bigO(1)$ memory. To prove that the proposed $\bigO(\log N)$ protocol can provide a higher  percentage of $\Gamma$-synchronized clocks, we simulated three scenarios where the proposed protocol is used. We compared the percentage of  $\Gamma$-synchronized clocks to the same scenarios but this time when the simple protocol is used. For all scenarios, the proposed protocol provides 70\% to 80\%  $\Gamma$-synchronized clocks while the simple protocol can only provide 20\% to 30\%  $\Gamma$-synchronized clocks. Our protocol improved the number of  $\Gamma$-synchronized clocks by at most 400\% during the same time span.

\bibliography{clock-sync}

\begin{thebibliography}{27}
\providecommand{\natexlab}[1]{#1}
\providecommand{\url}[1]{\texttt{#1}}
\expandafter\ifx\csname urlstyle\endcsname\relax
  \providecommand{\doi}[1]{doi: #1}\else
  \providecommand{\doi}{doi: \begingroup \urlstyle{rm}\Url}\fi

\bibitem[Aldaba(2008)]{aldaba08}
R.M. Aldaba.
\newblock \emph{Assessing Competition in {P}hilippine Markets}.
\newblock 2008.

\bibitem[{Ar reyouchi} et~al.(2013){Ar reyouchi}, Ghoumid, Ameziane, and
  Mrabet]{arreyouchi13}
E.{m}. {Ar reyouchi}, K.~Ghoumid, K.~Ameziane, and O.E. Mrabet.
\newblock Performance analysis of round trip time in narrowband {RF} networks
  for remote wireless communications.
\newblock \emph{International Journal of Computer Science and Information
  Technology}, 5\penalty0 (5):\penalty0 1--20, 2013.

\bibitem[Balboa et~al.(2010)Balboa, Prado, and Yap]{balboa10}
J.D. Balboa, F.L.E.~Del Prado, and J.T. Yap.
\newblock \emph{Achieving the {ASEAN} {E}conomic {C}ommunity 2015: {C}hallenges
  for the {P}hilippines}.
\newblock 2010.

\bibitem[Biaz and Vaidya(2003)]{biaz03}
S.~Biaz and N.~Vaidya.
\newblock Is the round-trip time correlated with the number of packets in
  flight?
\newblock In \emph{ACM SIGGCOMM Internet Measurement Conference}, 2003.

\bibitem[Castro and Pabico(2013{\natexlab{a}})]{castro13a}
F.E.V.G. Castro and J.P. Pabico.
\newblock A study on the effect of exit widths and crowd sizes in the formation
  of arch in clogged crowds.
\newblock \emph{Philippine Computing Journal}, 8\penalty0 (1):\penalty0 21--29,
  2013{\natexlab{a}}.
\newblock ISSN: 1908-1995.

\bibitem[Castro and Pabico(2013{\natexlab{b}})]{castro13b}
F.E.V.G. Castro and J.P. Pabico.
\newblock Microsimulations of arching, clogging, and bursty exit phenomena in
  crowd dynamics.
\newblock \emph{Philippine Information Technology Journal}, 6\penalty0
  (1):\penalty0 11--16, 2013{\natexlab{b}}.
\newblock ISSN: 2012-0761.

\bibitem[Congress(1978)]{bp8}
Philippine Congress.
\newblock {Section 6(b) of \bf Batas Pambansa Bilang 8: An Act Defining the
  Metric System and Its Units, Providing for Its Implementation and for Other
  Purposes}, 1978.

\bibitem[Costa et~al.(2008)Costa, Silva, Fedak, and Kelley]{costa08}
F.~Costa, L.~Silva, G.~Fedak, and I.~Kelley.
\newblock Optimizing the data distribution layer of {BOINC} with
  {B}it{T}orrent.
\newblock In \emph{Proceedings of the IEEE International Symposium on Parallel
  and Distributed Processing}, 2008.
\newblock doi:10.1109/IPDPS.2008.4536446.

\bibitem[Gotoh et~al.(2002)Gotoh, Imamura, and Kaneko]{gotoh02}
T.~Gotoh, K.~Imamura, and A.~Kaneko.
\newblock Improvement of {NTP} time offset under the asymmetric network with
  double packets method.
\newblock In \emph{Proceedings of the IEEE 2002 Conference on Precision
  Electromagnetic Measurements}, pages 448 -- 449, 2002.

\bibitem[Gruber and Holzer(2009)]{gruber09}
H.~Gruber and M.~Holzer.
\newblock Language operations with regular expressions of polynomial size.
\newblock \emph{Theoretical Computer Science}, 410\penalty0 (35):\penalty0
  3281--3289, 2009.

\bibitem[Hollweg and Wong(2009)]{hollweg09}
C.~Hollweg and H.M. Wong.
\newblock \emph{Measuring Regulatory Restrictions in Logistics Services}.
\newblock 2009.

\bibitem[Iglesias(2010)]{iglesias10}
G.~Iglesias.
\newblock \emph{{e}-Government Initiatives of Four {P}hilippine Cities}.
\newblock 2010.

\bibitem[Kornhauser et~al.(2007)Kornhauser, Rand, and Wilensky]{kornhauser07}
D.~Kornhauser, W.~Rand, and U.~Wilensky.
\newblock Visualization tools for agent-based modeling in {N}et{L}ogo.
\newblock In \emph{Proceedings of the AGENT}, 2007.

\bibitem[Marzullo(1984)]{marzullo84}
K.A. Marzullo.
\newblock \emph{Maintaining the Time in a Distributed System: An Example of a
  Loosely-Coupled Distributed Service}.
\newblock PhD thesis, Stanford University, February 1984.

\bibitem[Maslov(1973)]{maslov73}
A.N. Maslov.
\newblock Cyclic shift operation for languages.
\newblock \emph{Problems of Information Transmission}, 3:\penalty0 333--338,
  1973.

\bibitem[Murdoch(2006)]{murdoch06}
S.J. Murdoch.
\newblock Hot or not: {R}evealing hidden services by their clock skew.
\newblock In \emph{Proceedings of the 13th ACM Conference on Computer and
  Communications Security}, 2006.
\newblock Alexandria, VA, USA, 30 October -- 3 November.

\bibitem[Muscalagiu et~al.(2013)Muscalagiu, Popa, and Vidal]{muscalagiu13}
I.~Muscalagiu, H.E. Popa, and J.~Vidal.
\newblock Clustered computing with {N}et{L}ogo for the evaluation of
  asynchronous search techniques.
\newblock In \emph{Proceedings of 12th IEEE International Conference on
  Intelligent Software Methodologies, Tools and Techniques (SOMET 2013)}, pages
  115--120, 2013.

\bibitem[Mutzel and Weiskircher(2000)]{mutzel00}
P.~Mutzel and R.~Weiskircher.
\newblock Computing optimal embeddings for planar graphs.
\newblock In \emph{Proceedings of the 6th Annual International Conference on
  Computing and Combinatorics (COCOON 2000)}, pages 95--104. Springer-Verlag,
  2000.
\newblock Lecture Notes in Computer Science 1858; DOI:
  10.1007/3-540-44968-X\_10.

\bibitem[Niazi(2011)]{niazi11}
M.~Niazi.
\newblock Agent-based computing from multi-agent systems to agent-based models:
  {A} visual survey.
\newblock \emph{Scientometrics}, 89\penalty0 (2):\penalty0 479--499, 2011.
\newblock DOI: 10.1007/s11192-011-0468-9.

\bibitem[Oshiba(1972)]{oshiba72}
T.~Oshiba.
\newblock Closure property of the family of context-free languages under the
  cyclic shift operation.
\newblock \emph{Transactions of IECE}, 55\penalty0 (D):\penalty0 119--122,
  1972.

\bibitem[Pabico(2014)]{pabico2014}
J.P. Pabico.
\newblock Paths with jumps: {D}efinition, topology-preserving dynamics, and
  applications.
\newblock \emph{Asia Pacific Journal of Education, Arts and Sciences},
  1\penalty0 (2):\penalty0 61--69, 2014.
\newblock ISSN : 2362-8022.

\bibitem[Plagger and Wilson(1986)]{plagger86}
D.~Plagger and W.K. Wilson.
\newblock Time corrected, continuously updated clock, 1986.
\newblock US Patent 4,582,434 issued 15 April.

\bibitem[Pouwelse et~al.(2005)Pouwelse, Garbacki, Epema, and Sips]{pouwelse05}
J.~Pouwelse, P.~Garbacki, D.~Epema, and H.~Sips.
\newblock \emph{The {B}ittorrent {P2P} File-Sharing System: {M}easurements and
  Analysis}, volume 3640, pages 205--216.
\newblock Springer, 2005.
\newblock Lecture Notes in Computer Science.

\bibitem[Salamon(2011)]{salamon11}
T.~Salamon.
\newblock \emph{Design of Agent-Based Models : Developing Computer Simulations
  for a Better Understanding of Social Processes}.
\newblock Bruckner Publishing, 2011.
\newblock ISBN: 978-80-904661-1-1.

\bibitem[Sessini and Mahanti(2006)]{sessini06}
P.~Sessini and A.~Mahanti.
\newblock Observations on round-trip times of {TCP} connections.
\newblock In \emph{Proceedings of the 2006 International Symposium on
  Performance Evaluation of Computer and Telecommunication Systems
  (SPECTS'06)}, 2006.

\bibitem[Stevens(1998)]{stevens98}
W.R. Stevens.
\newblock \emph{UNIX Network Programming: Networking APIs: Sockets and XTI},
  volume~1.
\newblock Prentice Hall, 2nd edition, 1998.
\newblock ISBN: 0-13-490012-X.

\bibitem[Whitney(1932)]{whitney32}
H.~Whitney.
\newblock Congruent graphs and the connectivity of graphs.
\newblock \emph{American Journal of Mathematics}, 54\penalty0 (1):\penalty0
  150--168, 1932.
\newblock DOI: 10.2307/2371086.

\end{thebibliography}
\bibliographystyle{plainnat}
\end{document}